\documentclass[prd,preprint,superscriptaddress,amsmath,amssymb,nofootinbib]{revtex4}
\usepackage{graphicx}
\usepackage{dcolumn}
\usepackage{bm}
\usepackage{amssymb}
\usepackage{amsmath}
\usepackage{epsfig}    
\usepackage{color}
\usepackage{slashed}
\usepackage{hhline}
\def\be{\begin{equation}}
\def\ee{\end{equation}}
\newcommand{\bea}{\begin{eqnarray}}
\newcommand{\eea}{\end{eqnarray}}
\newcommand{\nn}{\nonumber}

\numberwithin{equation}{section}

\begin{document}

{\begin{flushright}{APCTP Pre2021-028, CTP-SCU/2021034}
\end{flushright}}

\title{Zee model in a modular $A_4$ symmetry}

\author{Takaaki Nomura}
\email{nomura@scu.edu.cn}
\affiliation{College of Physics, Sichuan University, Chengdu 610065, China}

\author{Hiroshi Okada}
\email{hiroshi3okada@htu.edu.cn}
\affiliation{Department of Physics, Henan Normal University, Xinxiang 453007, China}
\affiliation{Asia Pacific Center for Theoretical Physics, Pohang 37673, Republic of Korea}
\affiliation{Department of Physics, Pohang University of Science and Technology, Pohang 37673, Republic of Korea}

\author{Yong-hui Qi}
\email{yonghui.qi@apctp.org}
\affiliation{Asia Pacific Center for Theoretical Physics, Pohang 37673, Republic of Korea}

\pacs{}
\date{\today}

\begin{abstract}
We study a Zee model applying a modular $A_4$ symmetry in supersymmetry framework, where we construct our lepton model as minimum modular weight as possible. We show our predictions on phases and neutrino masses through numrical analysis.  We find a model that can fit in both normal and inverted hierarchy cases of neutrino mass.  The predictions of neutrino observables are shown where we obtain more constrained allowed region in the inverted hierarchy case.
 \end{abstract}

\maketitle

\section{Introduction}
Understanding lepton masses and mixings is very important issue to clarify flavor puzzles in high energy physics.
In particular, mechanism of generating tiny neutrino masses is not uniquely determined yet even though there exist many ideas such as canonical seesaw, inverse seesaw, linear seesaw, and radiative seesaw.
Although the first three mechanisms are originally expected to be higher scale such as grand unified theories that are far beyond the observed scale, the last one is supposed that the theory has to be measured in our current experiments, that is, TeV scale. In addition, the radiative seesaw models strongly correlate with particles that run inside the loop. Thus, such fields are sometimes good candidates to be measured by current experiment such as the Large Hadron Collider (LHC).
But one needs additional symmetries, such as $Z_2$ symmetry that is minimum realization~\cite{Ma:2006km}, to maintain the loop models apart from few exceptions.
These symmetries can be extended into non-Abelian flavor symmetries.
In this case, we have bonuses in addition to assuring the loop models.
The most attractive bonus is that we have predictions of masses, mixings and phases on lepton sector due to decreasing free parameters.

Several years ago, a new type of non-Abelian flavor symmetry has been proposed in refs.~\cite{Feruglio:2017spp, deAdelhartToorop:2011re}, originated from modular symmetry. After that, a lot of literature have been arisen using this framework to restrict Yukawa coupling structure for realizing predictive models.
For example, 
the modular $A_4$ flavor symmetry has been discussed in refs.~\cite{Feruglio:2017spp, Criado:2018thu, Kobayashi:2018scp, Okada:2018yrn, Nomura:2019jxj, Okada:2019uoy, deAnda:2018ecu, Novichkov:2018yse, Nomura:2019yft, Okada:2019mjf,Ding:2019zxk, Nomura:2019lnr,Kobayashi:2019xvz,Asaka:2019vev,Zhang:2019ngf, Gui-JunDing:2019wap,Kobayashi:2019gtp,Nomura:2019xsb, Wang:2019xbo,Okada:2020dmb,Okada:2020rjb, Behera:2020lpd, Behera:2020sfe, Nomura:2020opk, Nomura:2020cog, Asaka:2020tmo, Okada:2020ukr, Nagao:2020snm, Okada:2020brs, Yao:2020qyy, Chen:2021zty, Kashav:2021zir, Okada:2021qdf, deMedeirosVarzielas:2021pug, Nomura:2021yjb, Hutauruk:2020xtk, Ding:2021eva, Nagao:2021rio, king, Okada:2021aoi}, 
$S_3$  in refs.~\cite{Kobayashi:2018vbk, Kobayashi:2018wkl, Kobayashi:2019rzp, Okada:2019xqk, Mishra:2020gxg, Du:2020ylx}, 
$S_4$  in refs.~\cite{Penedo:2018nmg, Novichkov:2018ovf, Kobayashi:2019mna, King:2019vhv, Okada:2019lzv, Criado:2019tzk,
Wang:2019ovr, Zhao:2021jxg, King:2021fhl, Ding:2021zbg, Zhang:2021olk, gui-jun, Nomura:2021ewm}, 
$A_5$ in refs.~\cite{Novichkov:2018nkm, Ding:2019xna,Criado:2019tzk},
double covering of $A_4$  in refs.~\cite{Liu:2019khw, Chen:2020udk, Li:2021buv}, 
double covering of $S_4$  in refs.~\cite{Novichkov:2020eep, Liu:2020akv},   and
double covering of $A_5$  in refs.~\cite{Wang:2020lxk, Yao:2020zml, Wang:2021mkw, Behera:2021eut}.
Other types of modular symmetries have also been proposed to understand masses, mixings, and phases of the standard model (SM) in refs.~\cite{deMedeirosVarzielas:2019cyj, Kobayashi:2018bff,Kikuchi:2020nxn, Almumin:2021fbk, Ding:2021iqp, Feruglio:2021dte, Kikuchi:2021ogn, Novichkov:2021evw}.~\footnote{Here, we provide useful review references for beginners~\cite{Altarelli:2010gt, Ishimori:2010au, Ishimori:2012zz, Hernandez:2012ra, King:2013eh, King:2014nza, King:2017guk, Petcov:2017ggy}.}
Different applications to physics such as dark matter and origin of CP violation are found in refs.~\cite{Kobayashi:2021ajl, Nomura:2019jxj, Nomura:2019yft, Nomura:2019lnr, Okada:2019lzv, Baur:2019iai, Kobayashi:2019uyt, Novichkov:2019sqv,Baur:2019kwi, Kobayashi:2020hoc, Tanimoto:2021ehw}.
Mathematical study such as possible correction from K\"ahler potential, systematic analysis of the fixed points,
moduli stabilization are discussed in refs.~\cite{Chen:2019ewa, deMedeirosVarzielas:2020kji, Ishiguro:2020tmo, Abe:2020vmv}.

In this paper, we apply the modular $A_4$ flavor symmetry to Zee model that is firstly proposed as a radiative seesaw model in ref.~\cite{Zee:1980ai}.
{ In the original Zee model, an extra $SU(2)$ singlet charged scalar field and second Higgs doublet are introduced with softly-broken $Z_2$ symmetry. Due to the $Z_2$ symmetry, only one Higgs doublet has Yukawa interaction. In that case, it is not possible to fit neutrino data since the neutrino mass matrix does not have sufficient free parameters~\cite{Koide:2001xy,Frampton:2001eu,He:2003ih}.  
We can fit the neutrino data by removing the $Z_2$ symmetry adopting general Yukawa couplings of two Higgs doublets~\cite{AristizabalSierra:2006ri,He:2011hs,Herrero-Garcia:2017xdu}. In that case, however, the neutrino sector is less predictive and flavor changing neutral current (FCNC) is induced in both lepton and quark sectors. 
Then application of modular $A_4$ symmetry is interesting to get predictive neutrino sector and avoid quark FCNC. 
The model in this work is a modified Zee model that is considered in supersymmetry (SUSY) framework and introduce inert Higgs doublet fields, in addition to type-II two Higgs doublet sector, propagating inside loop diagram generating neutrino mass. The new Higgs doublets are inert ones due to modular flavor symmetry and only couples to leptons so that we can avoid flavor constraints in quark sector choosing quarks uncharged under modular $A_4$ symmetry; also it is easier to evade constraints from lepton flavor violations (LFVs) since we can make inert scalars arbitrary heavy. 
 Then, we explore lepton sector in the model and show our predictions via numerical analysis, where we restrict ourselves the models to be as minimum modular weight as possible.}

This paper is organized as follows.
In Sec.~II, we review our minimum model constructing the renormalizable valid Lagrangian and potential.
Then, we numerically fix the Higgs masses and mixings so that our analysis makes it simpler.
After discussing the charged-lepton sector, we formulate the neutrino sector. In Sec.~III, we analyze numerical analysis in the lepton sector and show some predictions on the allowed region of $\tau$, phases, and neutrino masses. 
We have conclusions and discussion in Sec.~IV.

\begin{center} 
\begin{table}[tbh!]
\begin{tabular}{|c||c|c|}\hline\hline  
  & \multicolumn{2}{c|}{Leptons}   \\ \hline \hline
& ~$[ \hat{L}_{L_e}, \hat{L}_{L_\mu}, \hat{L}_{L_\tau}]$~& ~$\hat{\ell}_R^c$~       \\\hline\hline 
$SU(2)_L$ & $\bm{2}$  & $\bm{1}$     \\\hline 
$U(1)_Y$   & $-\frac12$ & $1$       \\\hline
 $A_4$ & $[1,1'',1']$ & $3$         \\ \hline
$-k_I$ & $ [-k_e,-k_\mu,-k_\tau]$ & $0$    \\
\hline
\end{tabular}
\caption{Superfield contents and  their charge assignments in Zee model under $SU(2)_L\times U(1)_Y\times A_4$ where $\hat{\ell}_R^c\equiv
[\hat{e}^c_R,\hat{\mu}^c_R,\hat{\tau}^c_R]^T$ and $-k_I$ is the number of modular weight.}
\label{tab:1}
\end{table}
\end{center}

\section{Model}
\begin{table}[t!]
\begin{tabular}{|c||c|c|c|c|c|c|c|c|}
\hline\hline  
                   & ~$\hat{H}_u$~  & ~${\hat{H}_d}$~  & ~$\hat{\eta}_u$~  & ~${\hat{\eta}_d}$~  & ~$\hat{s}^-$~ &  ~$\hat{s}^+$~ \\\hline 
$SU(2)_L$ & $\bm{2}$   & $\bm{2}$ & $\bm{2}$   & $\bm{2}$ & $\bm{1}$ & $\bm{1}$  \\\hline 
$U(1)_Y$   & $\frac12$  & $-\frac12$  & $\frac12$  & $-\frac12$ & $-1$  &  $1$     \\\hline
$A_4$   & $\bm{1}$     & $\bm{1}$ & $\bm{1}$     & $\bm{1}$  & $\bm{1}$ & $\bm{1}$    \\\hline
$-k_I$   & $0$ & $0$  & $-2$   & $-2$  &  $-4$ &  $-4$    \\\hline
\end{tabular}
\caption{\small 
Charge assignments for scalar sector in Zee model under $SU(2)_L\times U(1)_Y$.}
\label{tab:2}
\end{table}

{ In this section, we show our setup of the model and show phenomenological formulas. We extend the original Zee model introducing inert Higgs doublets and imposing modular $A_4$ symmetry under SUSY framework. In the model only the inert Higgs doublets and charged singlets contribute to neutrino mass generation due to the modular $A_4$ symmetry relaxing phenomenological constraints such as flavor ones. Structure of Yukawa couplings for neutrino mass generation are restricted by the modular $A_4$ giving some prediction for neutrino observables. }

 \subsection{Model setup}
Here, we formulate our modified Zee model imposing the modular $A_4$ symmetry in supersymmetry (SUSY) framework.
At first, we assign $[1_{ -k_e},1''_{ -k_\mu},1'_{ -k_\tau}]$ for left-handed leptons $[\hat L_{L_e},\hat L_{L_\mu},\hat L_{L_\tau}]$ under the modular $A_4$ symmetry, where the lower indices represent the values of modular weight; in our notation the fields with "hat" indicate superfield.
Here, $L_{L_a}\equiv [\nu_{L_a}, \ell_{L_a}]^T$ ($a\equiv e,\mu,\tau$) is the SM lepton doublet as the fermion component of $\hat{L}_a$
and $\hat{\ell}_R^c\equiv
[\hat{e}^c_R,\hat{\mu}^c_R,\hat{\tau}^c_R]^T$ is assinged to be triplet under $A_4$ with zero modular weight.
The charge assignments for matter superfields are summarized in Table~\ref{tab:1}.
In scalar sector, we introduce two isospin doublet Higgs superfields $\{\hat{H}_u,\ \hat{H}_d \}$ with modular weight $0$, two isospin doublet inert superfields $\{\hat{\eta}_u, \hat{\eta}_d \}$ with modular weight $-2$ and  two singly-charged superfields $\hat{s}^\pm$ with modular weight $-4$ where $H_u$ and $H_d$ compose the Higgs sector in minimal supersymmetric standard model (MSSM) ; here scalar sector means R-parity even components of superfields are scalar field.
We assign non-zero modular weight for $\hat{\eta}_{u,d}$ so that they do not mix with $\hat{H}_{u,d}$, and modular weight of $\hat{s}^\pm$ is chosen to obtain $\hat{L} \hat{L} \hat{s}^+$ term.
The new charged scalar bosons from $\hat{\eta}_d$ and $\hat{s}^+$ are needed to 
induce physical singly-charged scalar bosons that run inside the loop diagram inducing neutrino mass in the Zee model mechanism. 
We write Higgs doublets $H_i(i=u,d)$ by
$H_{u}=[w^{+}_u, (v_u+h_u+i z_u)/\sqrt2]^T$ and $H_{d}=[ (v_d+h_d+i z_d)/\sqrt2, w^{-}_d]^T$ with $v_H\equiv \sqrt{v_u^2+v_d^2}=246$ GeV; $H_u(H_d) \in \hat{H}_{u(d)}$.
 The massless states from linear combinations of $w^{+}_i(z_i)$ are Nambu-Goldstone(NG) bosons eaten by the SM gauge vector bosons $W^+(Z)$. 
The assignments in scalar sector are summarized in Table~\ref{tab:2}.
Superpotential for lepton sector under these symmetries is given by 
\begin{align}
  w_\ell & = 
\alpha_\ell \hat L_{L_e} \hat{H}_d (y_1^{(k_e)} \hat{e}^c_R + y_2^{(k_e)} \hat{\tau}^c_R + y_3^{(k_e)} \hat{\mu}^c_R) \ [+ (\alpha_\ell \to \alpha'_\ell, y^{(k_e)}_i \to y'^{(k_e)}_i)] \nn \\
& +\beta_\ell \hat L_{L_\mu} \hat{H}_d (y_3^{(k_\mu)}  \hat{\tau}^c_R  + y_1^{(k_\mu)} \hat{\mu}^c_R + y_2^{(k_\mu)} \hat{e}^c_R) \ [+ (\beta_\ell \to \beta'_\ell, y^{(k_\mu)}_i \to y'^{(k_\mu)}_i)] \nn\\
&+\gamma_\ell \hat L_{L_\tau} \hat{H}_d (y_2^{(k_\tau)} \hat{\mu}^c_R + y_1^{(k_\tau)} \hat{\tau}^c_R + y_3^{(k_\tau)} \hat{e}^c_R) \ [+ (\gamma_\ell \to \gamma'_\ell, y^{(k_\tau)}_i \to y'^{(k_\tau)}_i)]\nn\\
&+ a_\ell \hat L_{L_e} \hat{\eta}_d (y_1^{(k_e+2)}  \hat{e}^c_R + y_2^{(k_e+2)} \hat{\tau}^c_R + y_3^{(k_e+2)} \hat{\mu}^c_R) \ [+ (a_\ell \to a'_\ell, y^{(k_e+2)}_i \to y'^{(k_e+2)}_i)] \nn \\
&+  b_\ell \hat L_{L_\mu} \hat{\eta}_d (y_3^{(k_\mu+2)} \hat{\tau}^c_R + y_1^{(k_\mu+2)} \hat{\mu}^c_R + y_2^{(k_\mu+2)} \hat{e}^c_R) \ [+ (b_\ell \to b'_\ell, y^{(k_\mu+2)}_i \to y'^{(k_e+2)}_i)]\nn\\
&+  c_\ell \hat L_{L_\tau} \hat{\eta}_d (y_2^{(k_\tau+2)} \hat {\mu}^c_R + y_1^{(k_\tau+2)} \hat{\tau}^c_R + y_3^{(k_\tau+2)} \hat{e}^c_R) \ [+ (c_\ell \to c'_\ell, y^{(k_\tau+2)}_i \to y'^{(k_e+2)}_i)] \nn\\
  & +a' Y^{(k_e+k_\mu+4)}_{1'} \hat L_{L_e} (i\sigma_2) \hat L_{L_\mu} \hat s^+
  +b' Y^{(k_e+k_\tau+4)}_{1''} \hat L_{L_e} (i\sigma_2) \hat L_{L_\tau} \hat s^+
  + c' Y^{(k_\mu+k_\tau+4)}_{1} \hat L_{L_\mu} (i\sigma_2) \hat L_{L_\tau} \hat s^+ ,
\label{yukawa}
\end{align}
where $\sigma_2$ is the second component of the Pauli matrix, $A_4$ triplet modular form with weight $k_I$ is written by $Y_3^{(k_I)} = [y_1^{(k_I)},y_2^{(k_I)},y_3^{(k_I)}]$ ($Y_{3'}^{(k_I)} = [y'^{(k_I)}_1,y'^{(k_I)}_2,y'^{(k_I)}_3]$ being the other $A_4$ triplet if exists) ,
terms with bracket $[\cdots]$ appear when there are multiple triplet modular form having the same modular weight, and we include singlet modular form into complex parameters as 
$a \equiv a' Y^{(k_e+k_\mu+4)}_{1'}$, $b \equiv b' Y^{(k_e+k_\tau+4)}_{1'}$ and $c \equiv c' Y^{(k_\mu+k_\tau+4)}_{1'}$.
~\footnote{See e.g. ref.~\cite{Okada:2021aoi} for their concrete forms of modular forms that are functions of modulus $\tau$.}
In this model quarks are chosen to be trivial singlet of $A_4$ with $0$ modular weight. Then quarks have Yukawa interactions only with $H_{u,d}$ and get their masses as in the MSSM case. 
In this work we do not discuss quark sector and focus on neutrino mass production.

We consider following three cases of $\{k_e, k_\mu, k_\tau\}$;
\begin{align}
& {\rm Case 1:} \ \{k_e, k_\mu, k_\tau \} = \{2,2,2 \}, \\
& {\rm Case 2:} \ \{k_e, k_\mu, k_\tau \} = \{2,2,4 \}, \\
& {\rm Case 3:} \ \{k_e, k_\mu, k_\tau \} = \{2,4,4 \}.
\end{align} 
In case 1, we only have one triplet associated with $Y^{(2,4)}_3$ and number of parameters in superpotential Eq.~\eqref{yukawa} is the minimum as $\{\alpha_\ell, \beta_\ell, \gamma_\ell, a_\ell, b_\ell, c_\ell, a, b, c \}$.
In cases 2 and 3, there are additional parameters corresponding to primed ones in Eq.~\eqref{yukawa} when weight 6 modular forms are considered; $c'_\ell$ and $\{b'_\ell, c'_\ell\}$ for case 2 and 3 respectively. 
Notice here that we have four real parameters; $\alpha_\ell, \beta_\ell, \gamma_\ell, c$ and  the other complex ones;
$\{a_\ell, b_\ell, c_\ell, a,b\}$ for case 1, $\{a_\ell, b_\ell, c_\ell, c'_\ell, a,b\}$ for case 2 and $\{a_\ell, b_\ell, c_\ell, b'_\ell, c'_\ell, a,b\}$ for case 3 after phase redefinition of fields without loss of generality in the Yukawa sector.
Also parameters $\{\alpha_\ell, \beta_\ell, \gamma_\ell \}$ are fixed when we fit masses of the charged-leptons.

For scalar sector we focus on the terms that are relevant for neutrino mass generation.
The non-trivially valid term is given by 
\begin{align}
  {\cal V} &=
\mu  (H^T_d \cdot \eta_d)s^+ + {\rm h.c.} \supset \frac{\mu v_d}{\sqrt{2}} \eta^-_d s^+
  ,\label{Eq:pot}
\end{align}
where $ \mu\equiv \mu_0 Y^{(6)}_1$ is a complex mass scale parameter contributing to the neutrino mass matrix.  
In this analysis we only consider mixing between $\eta^+_d$ and $s^+$ assuming mixings related to $\eta^\pm_u$ are small choosing parameters associated with terms $\hat \eta_d \hat \eta_u$ and $\hat H_u  \hat \eta_u \hat s^-$ to be small.
The charged scalars $\eta^\pm_d$ and $s^\pm$ are written by mass eigenstates,
\begin{equation}
\begin{pmatrix} h^\pm \\ H^\pm \end{pmatrix} = \begin{pmatrix} \cos \theta &  \sin \theta \\ - \sin \theta & \cos \theta \end{pmatrix} 
\begin{pmatrix} s^\pm \\ \eta^\pm_d \end{pmatrix},
\end{equation} 
where we take mixing angle $\theta$ to be free parameter instead of original $\mu$ in the potential determining $\theta$.
The mass eigenvalues are written by $m_{h^\pm}$ and $m_{H^\pm}$ that are also taken to be free parameters.
We do not discuss $H_u$ and $H_d$ sector considering it is the same as the MSSM. 
For simplicity $\tan \beta \equiv v_u/v_d$ is fixed to be $5$ in numerical analysis.

\subsection{Charged-lepton mass matrix}
After the spontaneous electroweak symmetry breaking by $H_{u,d}$ VEVs,
the charged-lepton mass matrix $(M_e)_{LR}$ is given by
\begin{align}
&M_e = \frac{v_d}{\sqrt2} \tilde M_e ,\nn\\
&\tilde M_e =
\begin{pmatrix}
|\alpha_\ell| & 0 & 0 \\ 
 0 &  |\beta_\ell| &0 \\ 
0 & 0 &  |\gamma_\ell| \\ 
\end{pmatrix}
\begin{pmatrix}
 y_1^{(k_e)} & y_3^{(k_e)} & y_2^{(k_e)} \\ 
 y_2^{(k_\mu)} & y_1^{(k_\mu)} & y_3^{(k_\mu)} \\ 
y_3^{(k_\tau)} & y_2^{(k_\tau)} & y_1^{(k_\tau)} \\ 
\end{pmatrix}.
 \label{massmat}
\end{align}
The charged-lepton mass matrix is diagonalized by bi-unitary mixing matrix as $D_e\equiv{\rm diag}(m_e,m_\mu,m_\tau)=V^\dag_{eL} M_e V_{eR}$.
To fit the mass eigenvalues of charged-leptons, we solve the following three relations in terms of $|\alpha_\ell|$, $|\beta_\ell|$ and $|\gamma_\ell|$;
\begin{align}
&{\rm Tr}[M_e M_e^\dag] = |m_e|^2 + |m_\mu|^2 + |m_\tau|^2,\\
&{\rm Det}[M_eM_e^\dag] = |m_e|^2  |m_\mu|^2  |m_\tau|^2,\\
&({\rm Tr}[M_eM_e ^\dag])^2 -{\rm Tr}[(M_e M_e^{\dag})^2] =2( |m_e|^2  |m_\nu|^2 + |m_\mu|^2  |m_\tau|^2+ |m_e|^2  |m_\tau|^2 ),
\end{align}
fixing the three free parameters in numerical analysis.

\subsection{Active neutrino mass matrix} 
\label{neut}
Now we write the renormalizable Lagrangian for the neutrino sector in terms of mass eigenstates of charged-leptons and singly -charged bosons, and it is is found as follows:
\begin{align}
-{\cal L}_\nu &= \overline{\ell_R} G \nu_L  (s_\theta h^- + c_\theta H^-) + \nu^T_L F \ell_L   (c_\theta h^+ - s_\theta H^+) 
 +    {\rm h.c.} ,\label{eq:neut}
\end{align}
where $s_\theta (c_\theta)$ is abbreviation of $\sin \theta (\cos \theta)$.
The coupling matrix is given by
\begin{align}
\tilde G &= V_{eR}^\dag\tilde  g, \quad \tilde F = \tilde f V_{eL},  \\ 
g& = 
\begin{pmatrix}
 y_1^{(k_e)} & y_3^{(k_e)} & y_2^{(k_e)} \\ 
 y_2^{(k_\mu)} & y_1^{(k_\mu)} & y_3^{(k_\mu)} \\ 
y_3^{(k_\tau)} & y_2^{(k_\tau)} & y_1^{(k_\tau)} \\ 
\end{pmatrix} 
\begin{pmatrix}
a_\ell & 0 & 0 \\ 
 0 &  b_\ell &0 \\ 
0 & 0 &  c_\ell \\ 
\end{pmatrix} 
+
\begin{pmatrix}
 y'^{(k_e)}_1 & y'^{(k_e)}_3 & y'^{(k_e)}_2 \\ 
 y'^{(k_\mu)}_2 & y'^{(k_\mu)}_1 & y'^{(k_\mu)}_3 \\ 
y'^{(k_\tau)}_3 & y'^{(k_\tau)}_2 & y'^{(k_\tau)}_1 \\ 
\end{pmatrix} 
\begin{pmatrix}
0 & 0 & 0 \\ 
 0 &  b'_\ell &0 \\ 
0 & 0 &  c'_\ell \\ 
\end{pmatrix} 
\nn \\
& = a_\ell \left( \begin{pmatrix}
 y_1^{(k_e)} & y_3^{(k_e)} & y_2^{(k_e)} \\ 
 y_2^{(k_\mu)} & y_1^{(k_\mu)} & y_3^{(k_\mu)} \\ 
y_3^{(k_\tau)} & y_2^{(k_\tau)} & y_1^{(k_\tau)} \\ 
\end{pmatrix} 
+
\begin{pmatrix}
 y'^{(k_e)}_1 & y'^{(k_e)}_3 & y'^{(k_e)}_2 \\ 
 y'^{(k_\mu)}_2 & y'^{(k_\mu)}_1 & y'^{(k_\mu)}_3 \\ 
y'^{(k_\tau)}_3 & y'^{(k_\tau)}_2 & y'^{(k_\tau)}_1 \\ 
\end{pmatrix} 
\begin{pmatrix}
0 & 0 & 0 \\ 
 0 &  \tilde b'_\ell &0 \\ 
0 & 0 &  \tilde c'_\ell \\ 
\end{pmatrix} \right)
\begin{pmatrix}
1 & 0 & 0 \\ 
 0 &  \tilde b_\ell &0 \\ 
0 & 0 &  \tilde c_\ell \\ 
\end{pmatrix}  \equiv a_\ell \tilde g
\\
f&= \begin{pmatrix}
0  &  a & b \\ 
-a & 0 & |c| \\ 
-b & -|c| & 0 \\ 
\end{pmatrix}
=|c|
\begin{pmatrix}
0  &  \epsilon' & \epsilon \\ 
-\epsilon' & 0 &1 \\ 
-\epsilon & -1 & 0 \\ 
\end{pmatrix}
\equiv |c| \tilde f
,
\end{align}
where  $\ell_R\equiv[e_R,\mu_R,\tau_R]^T$, $\tilde b_\ell = b_\ell/a_\ell$ and $\tilde c_\ell = c_\ell/a_\ell$, $\tilde b'_\ell = b'_\ell/b_\ell$ and $\tilde c'_\ell = c'_\ell/c_\ell$,  $\epsilon'\equiv a/|c|$ and  $\epsilon\equiv b/|c|$.
Note here that $b'_\ell =c'_\ell= 0$ for case 1and $b'_\ell =0$ for case 2.
%
Then, the neutrino mass matrix is given at one-loop level as follows:
\begin{align}
& (m_{\nu})_{ij}  \approx- \frac{a_\ell |c| s_\theta c_\theta}{(4 \pi)^2} \sum_{a=1} \left(\tilde  F_{ia} m_{\ell_a} \tilde G_{aj} + \tilde G^T_{ia} m_{\ell_a} \tilde F^T_{aj} \right) 
\ln\left(\frac{m^2_{h^\pm}}{m^2_{H^\pm}}\right).
\end{align}
Note that the loop function does not affect structure of  the neutrino mass matrix because of $m_{e,\mu,\tau} \ll m_{h^\pm_m}$.
In performing numerical analysis we define scaled neutrino mass matrix as $\tilde m_\nu \equiv \kappa m_\nu$ where $\kappa = a_\ell |c| s_\theta c_\theta$.
The scaled neutrino mass matrix $\tilde m_\nu$ is diagonalized by a unitary matrix $U_\nu$ as $U_\nu^T \tilde m_\nu U_\nu =\tilde D_\nu$ with $\tilde D_\nu = {\rm diag}[\tilde m_1,\tilde m_2,\tilde m_3]$.
Then, the Pontecorvo-Maki-Nakagawa-Sakata unitary matrix $U_{PMNS}$~\cite{Maki:1962mu} is defined by $V_{eL}^\dag U_\nu$. The observed atmospheric mass squared difference $\Delta m^2_{atm}$ is written by 
\begin{align}
{\rm NH}:\ \Delta m^2_{atm}= |\kappa|^2 (\tilde m^2_3 - \tilde m^2_1),\\
{\rm IH}:\ \Delta m^2_{atm}= |\kappa|^2 (\tilde m^2_2 - \tilde m^2_3),
\end{align}
where NH(IH) represents normal(inverted) hierarchy.
The solar mass squared difference $\Delta m^2_{sol}$ is given by
\begin{align}
\Delta m^2_{sol}= |\kappa|^2 (\tilde m^2_2 - \tilde m^2_1).
\end{align}
We also evaluate 
the effective mass for the neutrinoless double beta decay by
\begin{align}
m_{ee}=\left| m_{1} c^2_{12} c^2_{13}+m_{2} s^2_{12} c^2_{13}e^{i\alpha_{21}}
+m_{3} s^2_{13}e^{i(\alpha_{31}-2\delta_{CP})} \right|.
\end{align}
Its value is tested by experiments, e.g. KamLAND-Zen~\cite{KamLAND-Zen:2022tow}. 

\subsection{Lepton flavor violations}

The Yukawa interactions inducing neutrino masses also contribute to lepton flavor violations (LFVs) decay of charged leptons.
The relevant Lagrangian for LFVs is 
\begin{align}
- \mathcal{L} = \overline{\ell_i} (s_\theta g_{ia} P_L + c_\theta f_{ia} P_R) \nu_a h^- + \overline{\ell_i} (c_\theta g_{ia} P_L - s_\theta f_{ia} P_R) \nu_a H^- + h.c. \ ,
\end{align}
where $g_{ia} \equiv (G U_\nu)_{ia}$, $f_{ia} \equiv (U^T_\nu F)^\dag_{ia}$ and $\nu_a$ indicates neutrino in mass basis.
Here we focus on LFV charged lepton decays $\ell_i \to \ell_j \gamma$ since they provide stringent constraints~\footnote{Other flavor constraints are also discussed for Zee model in refs.~\cite{Smirnov:1996bv,Herrero-Garcia:2017xdu} such as deviation of $\nu$-$e$ scattering and LFV Higgs decays.}.
These decay processes are induced at one-loop level and we can calculate the branching ratio (BR) such that
\begin{align}
BR(\ell_i \to \ell_j \gamma) =  &\frac{48 \pi^3 \alpha_{\rm em} C_{ij}}{144 (4 \pi)^4G _F^2} 
\left( \left| \frac{a_{h ja} a_{hai}^\dag}{m_{h^\pm}^2} + \frac{a_{H ja} a_{Hai}^\dag}{m_{H^\pm}^2} \right|^2 
+ \left| \frac{b_{h ja} b_{hai}^\dag}{m_{h^\pm}^2} + \frac{b_{H ja} b_{Hai}^\dag}{m_{H^\pm}^2} \right|^2 \right) ,
\end{align}
where we ignored lepton mass, $\{ a_{hia}, a_{Hia}, b_{hia}, b_{Hia} \} \equiv \{s_\theta g_{ia}, c_\theta g_{ia}, c_\theta f_{ia}, -s_\theta f_{ia} \}$ 
and $\{C_{21}, C_{31}, C_{32} \} = \{1, 0.1784, 0.1736 \}$.
Currently experimental upper bounds of the BRs are given 
by~\cite{MEG:2016leq, BaBar:2009hkt,Renga:2018fpd}
\begin{subequations}
\label{eq:lfvs-cond}
\begin{align}
& {\rm BR}(\mu\to e\gamma)\lesssim 3.1\times10^{-13}, \\
& {\rm BR}(\tau\to e\gamma)\lesssim 3.3\times10^{-8}, \\
& {\rm BR}(\tau\to\mu\gamma)\lesssim 4.4\times10^{-8}.
\end{align}
\end{subequations}
In the numerical analysis we impose these constraints.

\section{Numerical analysis}
In this section we carry out numerical analysis in which we search for parameters that can fit the data from NuFit5.0~\cite{Esteban:2018azc}.
The theoretical complex parameters are randomly scanned within the following ranges:
\begin{align}
&[|\tilde b_\ell |, |\tilde c_\ell |, |\tilde b'_\ell |, | \tilde c'_\ell |]\in [10^{-3},1],\quad 
[|\epsilon|, |\epsilon'|]\in [10^{-3},1], \nn \\
& |s_\theta| \in [1, 10^{-5}], \quad \{m_{h^\pm}, m_{H^\pm} \} \in [100, 10^5] \ {\rm GeV}.
\end{align}
Also modulus $\tau$ is scanned in fundamental region.

As a result of numerical analysis we could not find parameter region fitting the data when $\{k_e, k_\mu, k_\tau \} = \{2,2,2 \}$ and
$\{2,2,4 \}$. 
\footnote{We have found the minimum $\Delta \chi^2$ are respectively 1446(NH) for model $ \{2,2,2 \}$ and 148(NH) for model $ \{2,2,4 \}$.
IH gives larger values of the minimum $\Delta \chi^2$ than the ones of NH. }
On the other hand we find parameter region to fit the data for $\{k_e, k_\mu, k_\tau \} = \{2,4,4 \}$. 
In the following we show the results in the case for both NH and IH cases.

\subsection{NH}

\begin{figure}[t]
  \includegraphics[width=77mm]{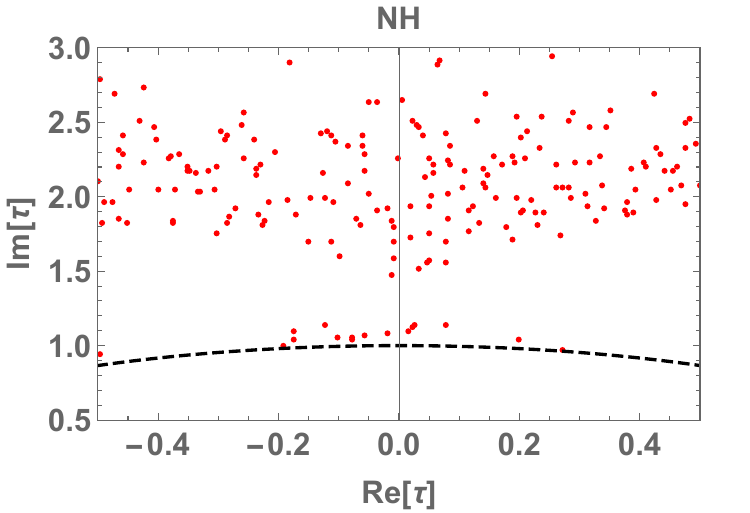}
  \caption{Allowed parameter points for real $\tau$ and imaginary $\tau$ in NH.}
  \label{fig:tau_nh}
\end{figure}

In Fig.~\ref{fig:tau_nh}, we show the allowed region of $\tau$, and find it around Im$(\tau) \sim 1$ and $1.5 \lesssim {\rm Im}(\tau) \lesssim 3$
  where Re$(\tau)$ is not constrained much but [$-0.2,  0.2 $] region is preferred for Im$(\tau) \sim 1$.

\begin{figure}[t]
  \includegraphics[width=77mm]{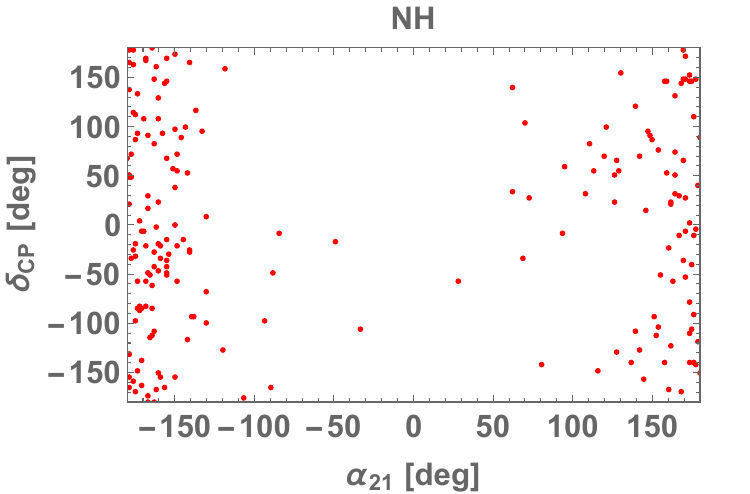}\quad
  \includegraphics[width=77mm]{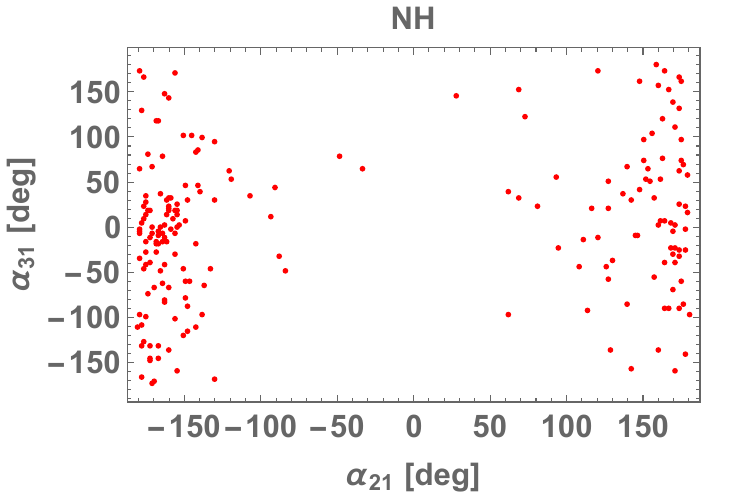}
  \caption{Allowed points for phases of $\alpha_{21}-\delta_{CP}$ in the left figure and  $\alpha_{21}-\alpha_{31}$ in the right figure in NH.}
  \label{fig:phases_nh}
\end{figure}

\begin{figure}[t]
  \includegraphics[width=77mm]{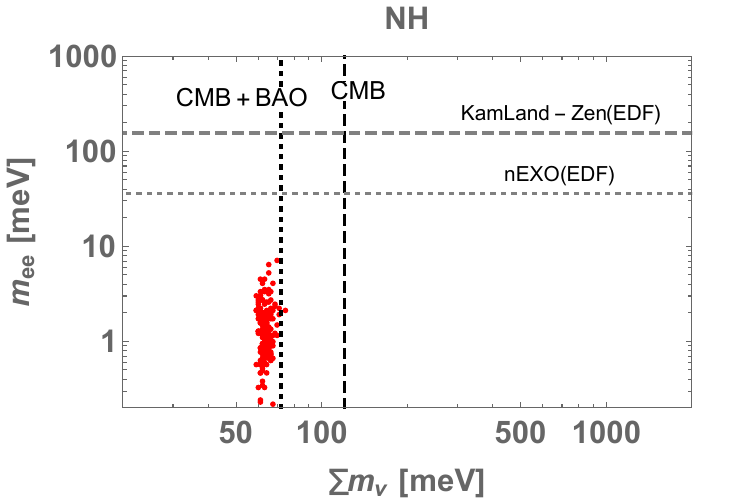}
    \includegraphics[width=77mm]{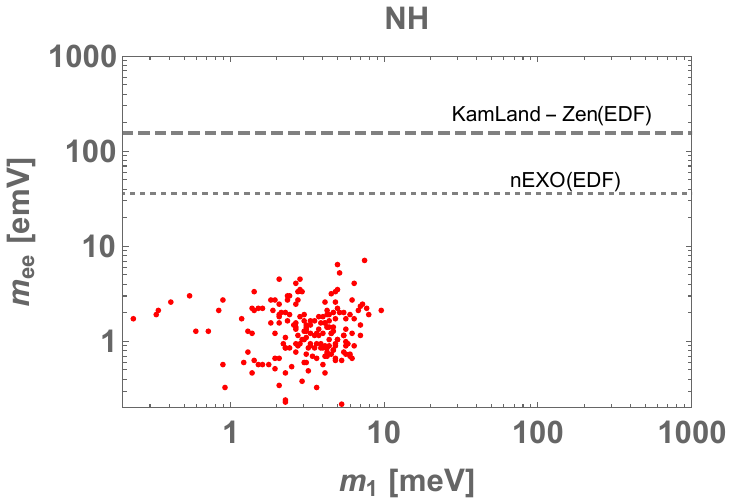}
  \caption{Allowed points for masses $\sum m_i$ and $\langle m_{ee}\rangle$  in the left and $m_1$  and $\langle m_{ee}\rangle$ in NH.}
  \label{fig:masses_nh}
\end{figure}

%
In Fig.~\ref{fig:phases_nh}, we show the allowed points for phases  of $\alpha_{21}-\delta_{CP}$ in the left figure and  $\alpha_{21}-\alpha_{31}$ in the right figure in the numerical analysis.
We find that $\alpha_{21}$ value is localized at around $\alpha_{21}=[\pm 150,  \pm180]$ [deg] and less points in the other region. 
On the other hand values of $\delta_{CP}$ and $\alpha_{31}$ do not have clear tendency and distributions are rather universal.
%
In left figure of Fig.~\ref{fig:masses_nh}, we show the allowed points for sum of masses $\sum m_i$ and $ m_{ee}$.
It displays that the allowed regions localized at $\sum m_i =[59 - 72]$ [meV]
and $\langle m_{ee}\rangle =[0.1 - 7]$ [meV]. It implies that there is large hierarchies among three neutrino mass eigenstates; $m_1<< m_2<<m_3$, because $\sum m_i$ is close to $\sqrt{\Delta m^2_{atm}}$. 
Also the value of $\sum m_i$ is compared with cosmological constraints from Planck~\cite{Planck:2018vyg} and Planck+DESI~\cite{DESI:2024mwx} where our values are below both the constraints.
The value of $m_{ee}$ is compared with the current constraint from KamLand-Zen~\cite{KamLAND-Zen:2022tow} and future prospect in nEXO~\cite{nEXO:2017nam} with energy-density functional (EDF) theory for nuclear matrix element. and we find our values are much smaller than these values. 
In right figure of Fig.~\ref{fig:masses_nh}, we also show the allowed points on $\{m_1, m_{ee} \}$. 
We find the predicted region of the lightest neutrino mass $m_1$ is around $[0.1, 10]$ meV.
{We also note that the LFV constraints exclude light $h^\pm$ and $H^\pm$ region. The constraints require $m_{h^\pm} \gtrsim 1000$ GeV and  $m_{H^\pm} \gtrsim 5000$ GeV.}

\subsection{IH}

\begin{figure}[t]
  \includegraphics[width=77mm]{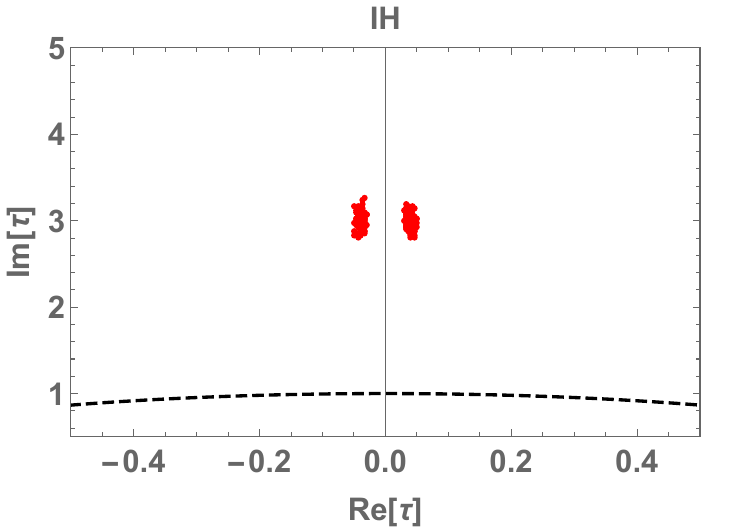}
  \caption{Allowed parameter points for real $\tau$ and imaginary $\tau$ in IH.}
  \label{fig:tau_ih}
\end{figure}

In Fig.~\ref{fig:tau_ih}, we show the allowed region of $\tau$, and find $\tau$ is localized at around ${\rm Im}(\tau) \in [2.7, 3.3]$ and  ${\rm Re}(\tau) \in [\pm 0.025, \pm 0.05]$ .

\begin{figure}[t]
  \includegraphics[width=77mm]{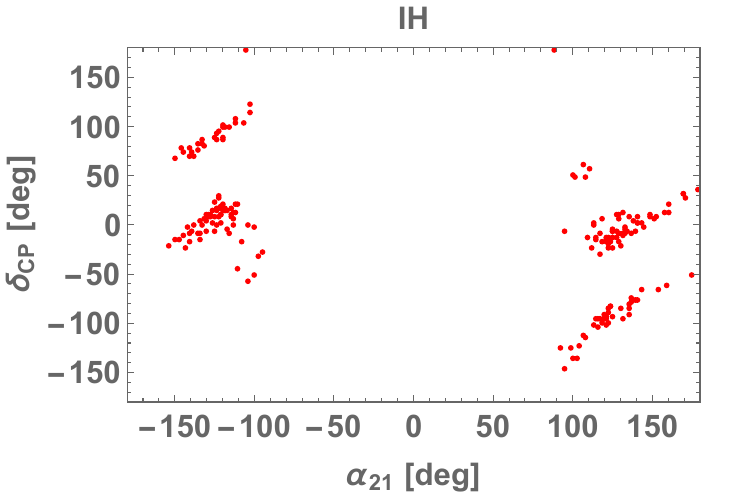}\quad
  \includegraphics[width=77mm]{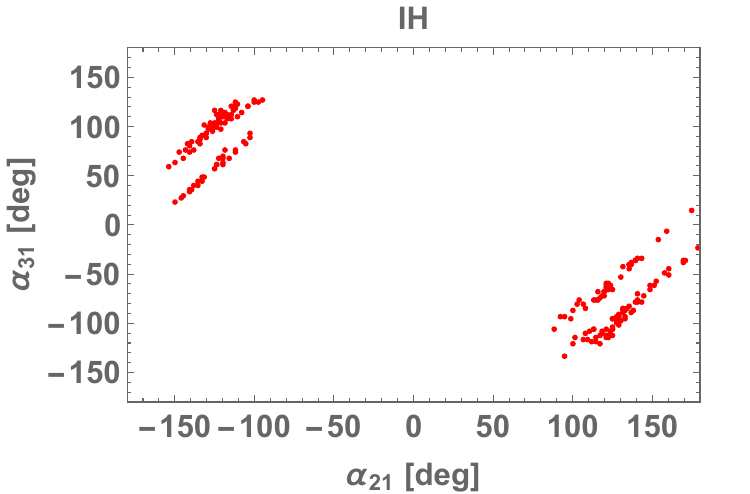}
  \caption{Allowed points for phases of $\alpha_{21}-\delta_{CP}$ in the left figure and  $\alpha_{21}-\alpha_{31}$ in the right figure in IH.}
  \label{fig:phases_ih}
\end{figure}

\begin{figure}[t]
  \includegraphics[width=77mm]{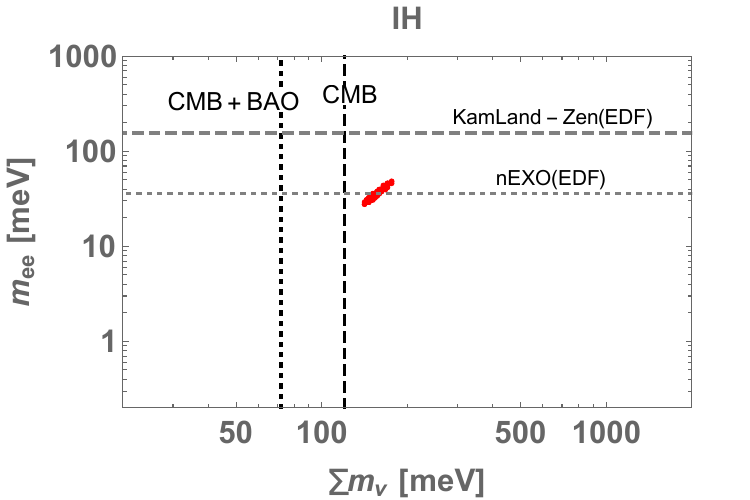}
    \includegraphics[width=77mm]{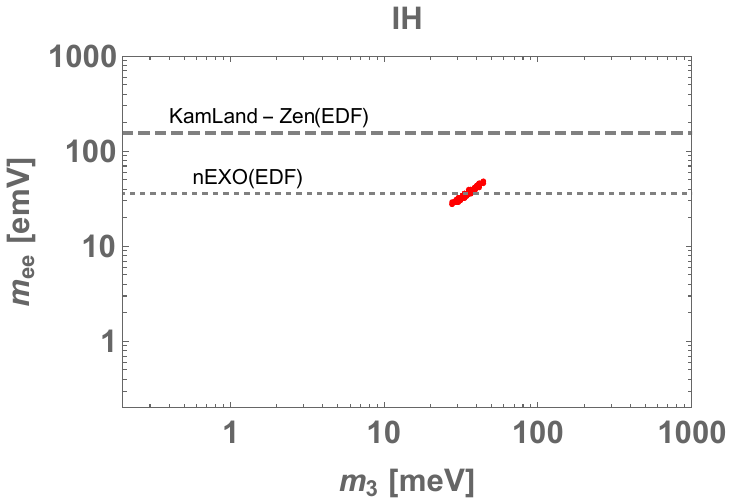}
  \caption{Allowed points for masses $\sum m_i$ and $\langle m_{ee}\rangle$  in the left and $m_3$  and $\langle m_{ee}\rangle$ in IH.}
  \label{fig:masses_ih}
\end{figure}

%
In Fig.~\ref{fig:phases_ih}, we show the allowed points for phases  of $\alpha_{21}-\delta_{CP}$ in the left figure and  $\alpha_{21}-\alpha_{31}$ in the right figure in the numerical analysis.
The figure demonstrates that the allowed regions of Majorana phases are localized around $\alpha_{21}=[(-160)-(-105),\ 90-180]$ [deg.]  $\alpha_{31}=[(-140)-20,\ 10-130]$ [deg.] and $\delta_{CP}=[30-100,\ 260-330]$ [deg]. and there are several localized region of $\delta_{CP}$.
%
In left figure of Fig.~\ref{fig:masses_ih}, we show the allowed points for sum of masses $\sum m_i$ and $ m_{ee}$.
It displays that the allowed regions are localized at $\sum m_i =[140 - 175]$ [meV]
and $\langle m_{ee}\rangle =[28 - 48]$ [meV].
In this case $\sum m_i $ is above the Planck constraint and $m_{ee}$ would be tested in future neutrinoless double beta decay experiments.
In right figure of Fig.~\ref{fig:masses_ih}, we also show the allowed points on $\{m_3, m_{ee} \}$. 
We find the predicted region of the lightest neutrino mass $m_3$ is around $[28 - 44]$ meV.
{We also note that the LHV constraints exclude light $h^\pm$ and $H^\pm$ region. The constraints require $m_{h^\pm} \gtrsim 1.8 \times 10^4$ GeV and  $m_{H^\pm} \gtrsim 5 \times 10^{4}$ GeV. }

\subsection{Comment on collider physics}
Here we briefly discuss signature of our model at collider such as the LHC experiments. 
In addition to the SM-like Higgs boson, In this model, there are two charged scalar bosons, heavy neutral CP-even boson and CP-odd boson which can be produced at the collider similar to two Higgs doublet model.
These scalar bosons decay into SM fermions via Yukawa couplings.
Interestingly we would have lepton flavor violating mode such as $\eta_d(h_d) \to \ell^+ \ell'^-$ since both $\eta_d$ and $h_d$ have Yukawa interactions with leptons.
Analysis of collider signature is beyond the scope of this work and it is left in future work.

\section{Conclusions and discussions}
We have investigated the Zee model applying a modular $A_4$ symmetry in supersymmetric framework, where we have constructed our model  as minimum modular weight as possible.
In the model scalar doublet inside a loop diagram generating neutrino mass is inert one due to modular symmetry and flavor constraints can be easily avoided compared to original Zee model.
We then searched for parameter region that can fit neutrino data in both NH and IH cases while evading lepton flavor violating constraints.
Then we have found one case of modular weight assignments for lepton doublet; $\{2,4,4\}$ model, can fit the data.

For NH case, we have obtained allowed $\tau$ region around Im$(\tau) \sim 1$ and $1.5 \lesssim {\rm Im}(\tau) \lesssim 3$. 
We have also found $\alpha_{21}=[\pm 150,  \pm180]$ [deg] is preferred region, $\sum m_i =[59 - 72]$ [meV] and $\langle m_{ee}\rangle =[0.1 - 7]$ [meV].
These values are below the current constraints.

For IH case, we have obtained more restricted region of $\tau$ around ${\rm Im}(\tau) \in [2.7, 3.3]$ and  ${\rm Re}(\tau) \in [\pm 0.025, \pm 0.05]$. 
The allowed regions of neutrino mass observables are also localized at $\sum m_i =[140 - 175]$ [meV],
 $ m_{ee} =[28 - 48]$ [meV] and $m_3 = [28-44]$ meV.
In this case $\sum m_i $ is above the Planck constraint and $m_{ee}$ would be well-tested in future neutrinoless double beta decay experiments.

\section*{Acknowledgments}
\vspace{0.5cm}
{
This research was supported by an appointment to the JRG Program at the APCTP through the Science and Technology Promotion Fund and Lottery Fund of the Korean Government. This was also supported by the Korean Local Governments - Gyeongsangbuk-do Province and Pohang City (H.O.). 
Y. O. was supported from European Regional Development Fund-Project Engineering Applications of Microworld
Physics (No.CZ.02.1.01/0.0/0.0/16\_019/0000766)}

\end{document}